\documentclass[a4paper,12pt]{article}
\usepackage[utf8]{inputenc}
\usepackage{graphicx}
\usepackage{amsthm,amsmath}
\usepackage[FIGTOPCAP]{subfigure}

\newcommand{\etal}{\emph{et al.}}

\title{Radial percolation reveals that Cancer Stem  Cells are trapped in the core of tumorspheres.}
\author{Lucas Barberis}

\begin{document}
\maketitle
\begin{abstract}

Using geometrical arguments it is shown that Cancer Stem Cells (CSC) must be confined inside solid tumors under natural situations. Aided by an agent-based model and percolation theory, the probability for a CSC to be at the border of a colony is estimated. This probability is estimated as a function of the CSCs self-renewal probability $p_s$, i.e. the chance that an CSC become non-differentiate after mitosis. In the most common situations, $p_s$ is small and CSCs are mostly destined to produce differentiated cells at a very low rate. The results presented here show that CSCs form a small core in the center of the cancer cell colony, becoming quiescent due to the lack of space to proliferate, which stabilizes their population size. This result provides a simple explanation for the CSC niche size that dispenses with the need of quorum sensing or other proposed signalling mechanisms. It is also supports the hypothesis that metastases are likely to start at the very beginning of the tumoral development.

\end{abstract}

\section{Introduction}
Cancer Stem Cells (CSCs) cells responsible for driving tumor growth   through their abilitiy to make copies of themselves (self-renewal) and to differentiate into cells with a more specific function \cite{Batlle2017}. The Differentiated Cancer Cells (DCCs) maintain a limited potential to proliferate but may only generate cells of their specific lineage \cite{LaPorta2012}.

Like the tissues from which they arise, solid tumors are composed of a heterogeneous population of cells; many properties of normal stem cells are shared by at least a subset of cancer cells \cite{Lobo2007,Stingl2007}. In many tissues, normal stem cells must be able to migrate to different regions of an organ where they give rise to the specifically differentiated cells requested by the organism. These features are reminiscent of invas ion and immortlity, two hallmark properties of cancer cells \cite{Shimono2009,Hanahan2000}. In fact, new therapeutic paradigms, based on the concept that destroying or incapacitating CSCs would be an efficient method of cancer containment and control, are at the focus of current research. Resistance to chemo/radio therapies gives to CSCs high chances of survival to forming new tumors even after treatment .\cite{Jagust2019}

An experimental biological model to study CSCs features is called a \emph{tumorsphere assay}. A tumorsphere  is a clonal aggregate of cancer cells grown \emph{in vitro} from a single cell.
DCCs  can not generate a tumorsphere because they experimentally lack the capability to form compact long-term aggregates. As a consequence, the current experimental convention   to define a CSC from a functional point of view is through the capability of a single cell, the \emph{seed}, to grow a more or less spherical aggregate in a gel suspension. That why it is said that CSCs ``drive" tumor progression. 

In an experimental assay, by measuring the time evolution of the number of cells in a tumorsphere, it is possible to determine a proliferation rate $r$ consistent with the \emph{Population Doubling Time} (PDT) of the total population. Unfortunately, such quantity cannot discriminate between the growth rates of CSCs and their differentiated counterparts. Furthermore, measuring the  PDT of the CSCs is experimentally very complex \cite{Waisman2019}. Understanding $r$ is crucial for the mathematical modelling in a system where, unlike the usual, the offpring belong to a different population than their parents. Indeed, a CSC division could have three possible outcomes: Stem cell replication (self-renewal), assymetric differentiation and symmetric differentiation. To mathematically model such a feature, we can assume that the three outcomes will occur with probabilities $p_s$, $p_a$ and $p_d$ respectively.  From the point of view of the populations, the last two possibilities correspond to the birth of a new DCC, keeping (in the asymmetric division) or losing (in the symmetric case) the parent cell its stemness. As an example, writing the corresponding  population dynamics differential equations, Benítez \etal \, showed that a CSC will give birth to another CSC at a rate $r p_s$  which is estimated by fitting experimental data with their  mathematical model \cite{Benitez2019, Benitez2020}. The probability of self-replication $p_s$ seems to be small in homeostasis and in most common culture media, then $rp_s$ will be also small leading to \emph{quiescence}, a main feature of CSCs. Nevertheless, CSCs can be experimentally forced to abandon its quiescent state by using specific growth factors that inhibit differentiation \cite{Wang2016, Chen2016} or by restricting the oxygen concentration as discussed below. In these situations, $p_s$ is close to one and tumorspheres will contain a high fraction of CSCs. 

Metastasis, the invasion process in which cancer cells leave a tumor to form a new colony at another location, it is an intriguing feature of cancer disease. Because CSCs are the seeds of tumors, the sources of metastatic tumors must come from a pre-existent  \emph{primary} tumor and might be located near its surface in order to detach, migrate and finally invade another place into the organism. 

The distribution of the CSCs in a primary tumor is a key feature for the  understanding of metastasis, cell proliferation and drig resistance. Curiously, CSCs seem to be in the inner core of glioblastomas, according to a three concentric layers model proposed by Persano \etal; menawhile cells in the periphery of the tumor show a more differentiated phenotype that is highly sensitive to Temozolomide, a drug for cancer treatment \cite{Persano2011}. Interestingly, these authors demonstrate that CSCs proliferate better under hypoxia conditions. Furthermore, Li \etal\,  report that hypoxia plays an important role in de-differentiation of cells \cite{Li2013}. These results state that an hypoxic environment will increase the scores of CSC. But, if CSCs are located in the core of a tumor under hypoxic conditions, what are the chances they will reach the periphery in order to detach and undergo metastasis?

To address such a question we use, without loss of generality, a two-dimensional model with which it is easier to visualize, characterize and understand the key geometrical features of the CSC distribution. Besides, it is computationally cheaper than simulate the whole tumorsphere. To keep this detail in mind, we will refer to our simulated models as \emph{colonies}. 

The aim of this work is to simulate the colony growth by means of an Agent Based Model (ABM) that mimics basic features of CSC proliferation with emphasis on its geometrical properties. Multiple realizations allow us to estimate the fraction of CSCs at the periphery of the colony, showing that it is really small for large long-lived colonies.  We also use some elements of percolation theory to help to interpret and quantify the simulation results. We then report a transition to percolation which depends on the self-renewal probability $p_s$ of the CSCs population. Finally, we conclude that that our results lead to a simple explanation for the CSCs niche size and support the hypothesis that the metastatic process must start at the very beginning of the tumoral development.

\section{Simulation of two-dimensional colonies.}

Experimental monoclonal colonies start with a CSC in a suitable culture medium. This cell and its daughters divide more or less at a constant rate forming a colony with an approximately circular shape. The reader interested in examples may take a look to \cite{Zhang2015,Shankar2011,Schneider2008} for further reference. 

In our computational simulations we use the same principle. A cell is a circle-shaped mathematical agent belonging to a particular class. In our simulations, these classes are: Active CSCs, active DCCs, quiescent CSCs and quiescent DCCs. Thus, we first seed a parent CSC and ask it to duplicate. The new cell will randomly search for empty space near to its parent. Then, each active cell will be asked, in a random order, to pop-up a new cell in its neighboring empty space at a random position.  If there is no room for the new cell because its parent is surrounded by another cells, the active cell does not duplicate and changes to its respective quiescent class. If there is room, the parent cell creates another cell in the empty space. After a successful division, if the  parent cell is a CSC, it replicates with a probability $p_s$, which means that the new cell will be a CSC, otherwise a DCC is created. Finally, if the parent and the new cells are different, there is a probability $1/2$ that the parent will move leaving the new cell in the original place. Naturally, a parent DCC will also create another DCC if there is enough space.  Once all the active cells are requested to divide, independently of the success of the attempt, we say that a one day-long time step is performed. This implies, without loss of generality, that $r\simeq 1$day$^{-1}$, a reasonably growth rate that will aid our intuition. The videos in the supplementary information, were examples with ten realizations each illustrate the process.  
        
\begin{figure}[h!]
\subfigure[]{
\centering
\includegraphics[width=0.45 \linewidth]{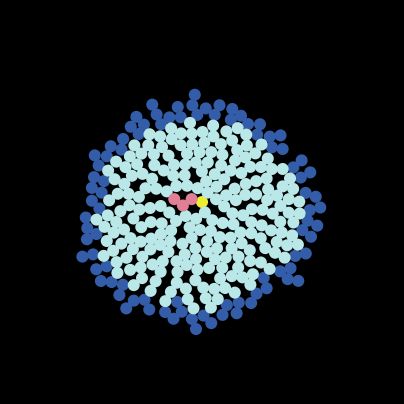}
}
\hfill
\subfigure[]{
\centering
\includegraphics[width=0.45 \linewidth]{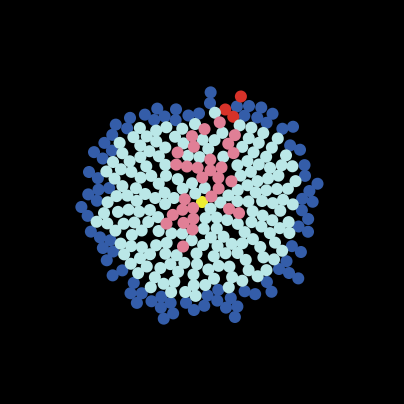}
}
\caption{Two realizations of the simulation for $p_s=0.5$ after 15 days. (a) There are no CSCs at the perifery; all of them, colored in pink, became quiescent. (b) The active CSCs at the border are colored in red. Note that some branches fail to percolate. DCCs are depicted in cyan (quiescent) and blue (active). The the seed is represented in yellow.  } \label{f 2d examples}
\end{figure}

Typical outcomes of the described  process are shown in Fig. \ref{f 2d examples} and video v1\_0.5.mp4.  In these examples we  set $p_s=0.5$ and start with a CSC seed depicted in yellow.  After running the simulation for a period equivalent to two weeks (15 time steps), a relatively long period for a biological experiment, we obtain two  possible outcomes of the realizations: (a) there are a few quiescent CSCs trapped into the center of the colony or (b) there are active CSCs at the border of the colony. Indeed, we define the \emph{border} of the colony as the rim formed by the set of  active cells. To better track the subpopulations present in the colony, the active cells are colored in red for CSC and blue for DCC. Also, the quiescent cells are colored in pink for CSC and in cyan for DCC. Recall that the yellow dot does not always be at the center of the colony because the seed has a chance of exchanging its place with a new DCC. 
In panel (b) we recognize a path, paved by CSCs, that joins the center of the colony with its border. Also, some frustrated branches of this CSCs path, which die in the quiescent core, appear indicating a great variability. Such a path resembles clusters percolation in porous media, lattices or networks and we will discuss them  in the next section.       

\begin{figure}[h!t]
\subfigure[$p_s=0.2$]{
\centering
\includegraphics[width=0.45 \linewidth]{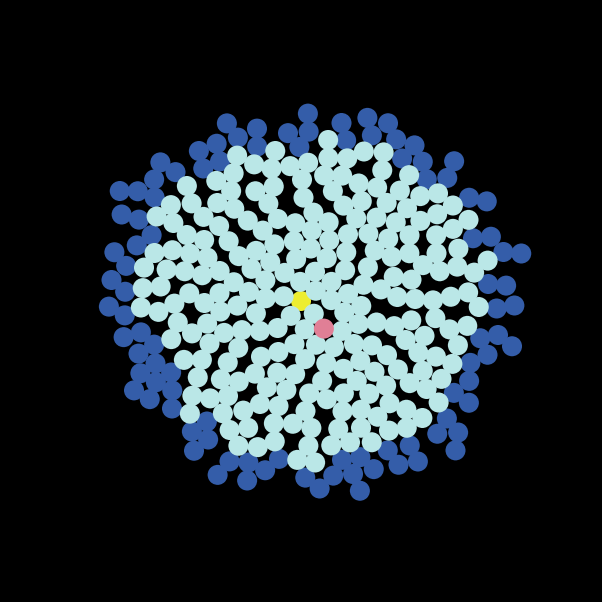}
}
\subfigure[$p_s=0.95$]{
\centering
\includegraphics[width=0.45 \linewidth]{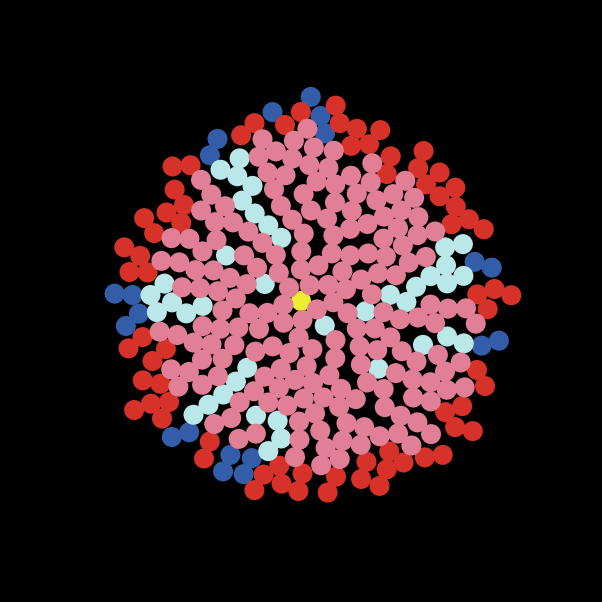}
}

\caption{Two realizations of the simulation at low (a) and   high (b) self-replication rates after 15 days. In (a) the CSC are quickly surrounded by DCC becoming quiescent (pink). In (b) there are so many CSC allowng percolation to most of the colony perimeter. The seed  is colored in yellow. } \label{f 2d examples2}
\end{figure}

To sharpen our intuition on what percolation means in this system we present, in Fig. \ref{f 2d examples2} and videos v2\_0.2.mp4 and v3\_0.95.mp4, two more examples at different self-replication probabilities $p_s$. In panel (a) and video v2\_0.2.mp4,  $p_s=0.2$ is so small that the CSCs are quickly surrounded by DCCs  and become quiescent. This is the most common situation in a regular culture medium where the CSCs number is low and constant along time. On the other hand, in panel (b) and video v3\_0.95.mp4, $p_s=0.95$ leads to a large CSC population that invades almost the whole system. The addition of stem cell maintenance factors, such as EGF or bFGF, to the culture medium is an example of this case. These limiting situations were previously studied both experimentally \cite{Wang2016, Chen2018} and mathematically \cite{Benitez2019,Benitez2020} focusing on technical aspects of the assay in the first and on recovering the CSC fraction in the second. 
 
Curiously, in the examples shown in Figs. (1) and (2), there is an active rim similar to the one surrounding a quiescent core, attributed to lack of oxygen/nutrient, in  the classical multicelular spheroids work of Freyer and Shuterland \cite{Freyer1986}. This rim formation was mathematically studied by several authors \cite{Condat2006, Menchon2008,Barberis2012} but for multicellular (not grown from a single cell) spheroids. In the present monoclonal case, geometry seems to be reason enough to put cells in the inner core in a quiescent state and to develop a rim, just two cells thick,  independently of any diffusion process.  

\begin{figure}[h!]
\centering
\includegraphics[width= \linewidth]{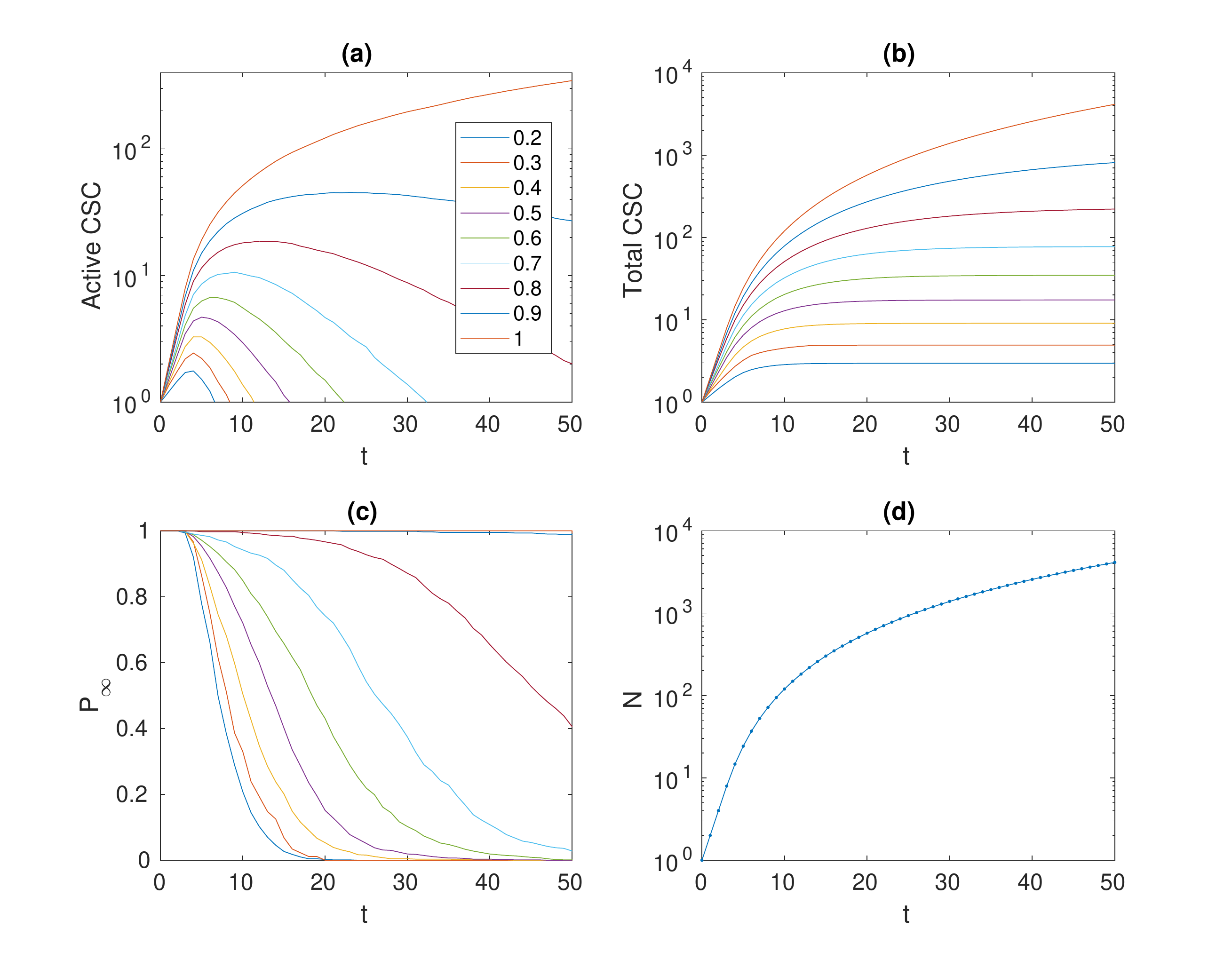}
\caption{Time evolutions of quantities averaed over a thousand realizations for values of $p_s \in [0.2 .. 1.0]$. (a) Number of CSCs at the border. (b) The total number of CSCs reaches a constant value that defines the niche. (c) The probability for a CSC to be at the border. (d) System size as function of time. } \label{f funciones de t}
\end{figure}

To statistically study the percolation properties of the system, we carry on  extensive simulations for the colonies: Each data point obtained is the result of averaging over 1000 realizations. We simulated the time evolution of the colony growth for many values of $p_s \in [0.2,\, 1.0]$. The results are summarized in Fig. \ref{f funciones de t}. Panel (a) shows the number of CSCs at the border, which increases to a maximum given by the time when they are overwhelmed by the DCC population. Beyond this point, more and more CSCs become quiescent until no more active CSCs left. This phenomenon disappears for $p_s$ barely lower than 1, when almost all cells in the border are CSCs. In panel (b) the time evolution of total CSC population is depicted showing that, after some transient, the CSC population stops growing and remains constant. This fixed number of CSC is usually called the \emph{CSC' niche} and was mathematically studied in \cite{Benitez2019}. As is intuitively expected, the probability of a CSC to be at the border  falls as $p_s$ is increased, as shown in panel (c). The relationship between the simulation time and the system size is shown in panel (d); due to its geometrical nature it is independent of $p_s$, the self-replication probability.

\section{Percolation theory}

Here we use percolation theory to estimate the probability of finding  a CSC at the border. We are looking for a purely geometrical feature of the colony growth. In particular, we assume that the cells in the colony can be mapped onto the nodes of a network with the seed at its  center. Then, each cell is connected with its nearest neighbors, whose number, in two dimension, can not exceed six. As shown Fig. \ref{f 2d examples}(b), the CSCs (quiescent and active) will form  \emph{paths} that radially expand from the center to the edge of the colony. Without loss of generality, we may assume that those paths are formed by the connected nodes that have had a CSC at any  moment during the culture time because some  ``holes", DCCs between two CSCs, could arise in the path due to the exchangig probability mentioned before. In the network a node can be occupied by a CSC with a probability $p_s$ or by a DCC otherwise, whenever it is connected to an occupied node (isolated nodes are forbidden). Thus, there must be a threshold value $p_c$ of $p_s$ below which it is impossible to make a continuous path of nodes occupied by CSCs that extends from the seed to the border of the tumor. Such value is called the \emph{critical value} or \emph{percolation threshold} and defines a \emph{percolation phase transition}\cite{Yonezawa1989}. Thus, when the border is reached by a CSC path, we will say that such a path \emph{percolates}.

Formally, a percolating path is mathematically defined as a set of connected nodes that expands infinitely over an infinite network for a value of ps large enough\cite{Christensen2005}. To define percolation in our model we note that, for a CSC to yield another CSC, two conditions are required: First the node that belongs to a CSC must be connected to an empty node.\footnote{In simulations this is equivalent to finding space to proliferate.} Second, the outcome of the mitosis must be another CSC. In the process of building up the path, it is important to note that several neighbouring nodes could be already occupied. Thus, it will be useful to define the  \emph{average {\bf empty} neighbour number} $z$as follows: Imagine that we perform a random walk starting at the seed node along an infinite path where it is {\bf forbidden to step back}. At each step there are $z$ branches of the path but only $p_sz$ will be accessible on the average. To continue walking along the path, there must be at least one node to walk to, that is, $p_s z \geq 1$. Therefore, the \emph{critical probability} for the transition to percolation is    

\begin{equation}\label{e 1/z}
p_c=\frac{1}{z},
\end{equation}  

\noindent which depends on the available neighbour number. \footnote{It is not universal because it depends on the details of the network.} For $z=1$ we obtain $p_c=1$, the 1D critical occupation of a chain of nodes. For $z>1$ the critical occupation is lower than 1. 

\begin{figure}[h!]
\centering
\subfigure[]{
\centering
\includegraphics[height= 0.4 \linewidth]{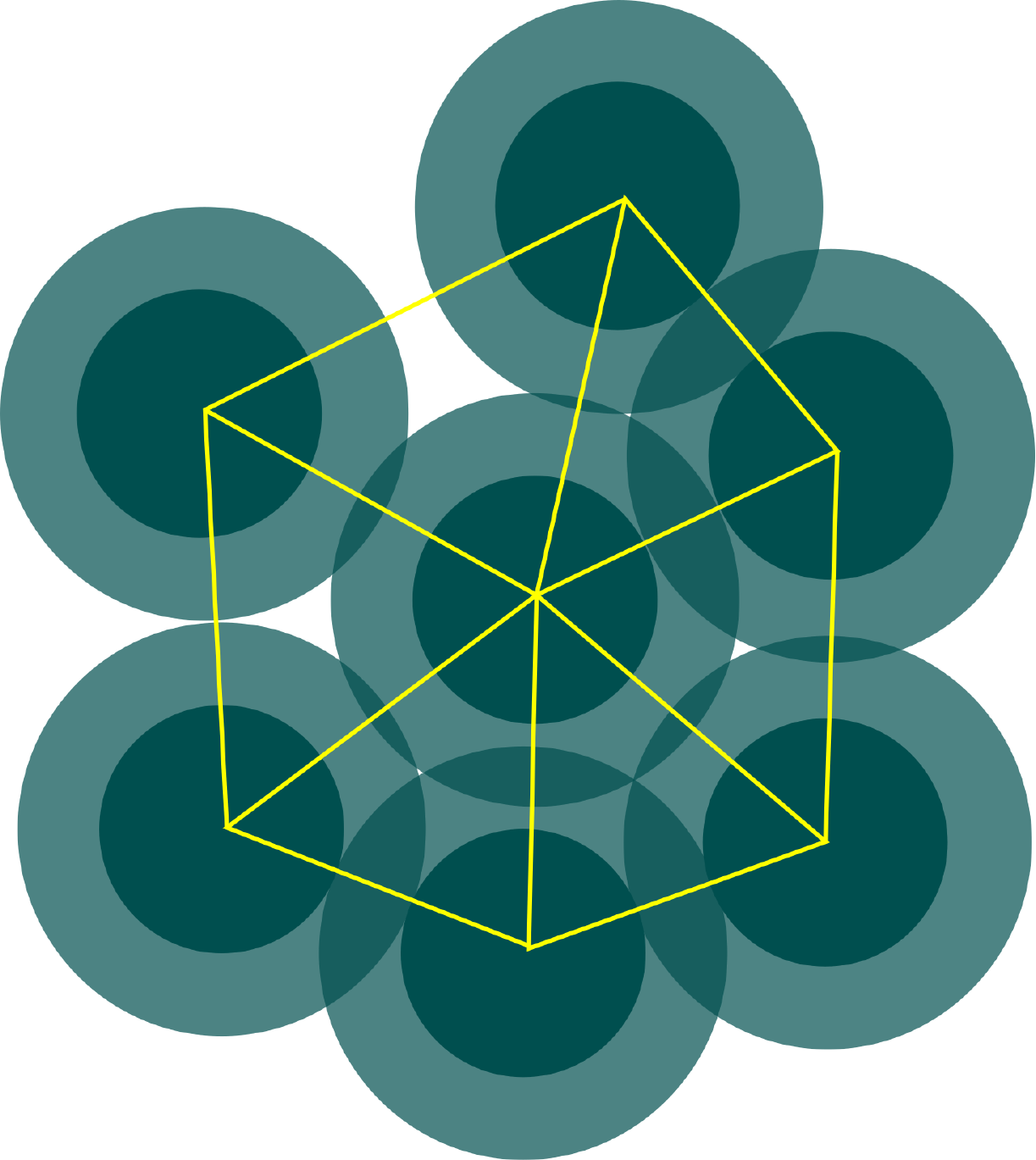}
}
\subfigure[]{
\centering
\includegraphics[height= 0.4 \linewidth]{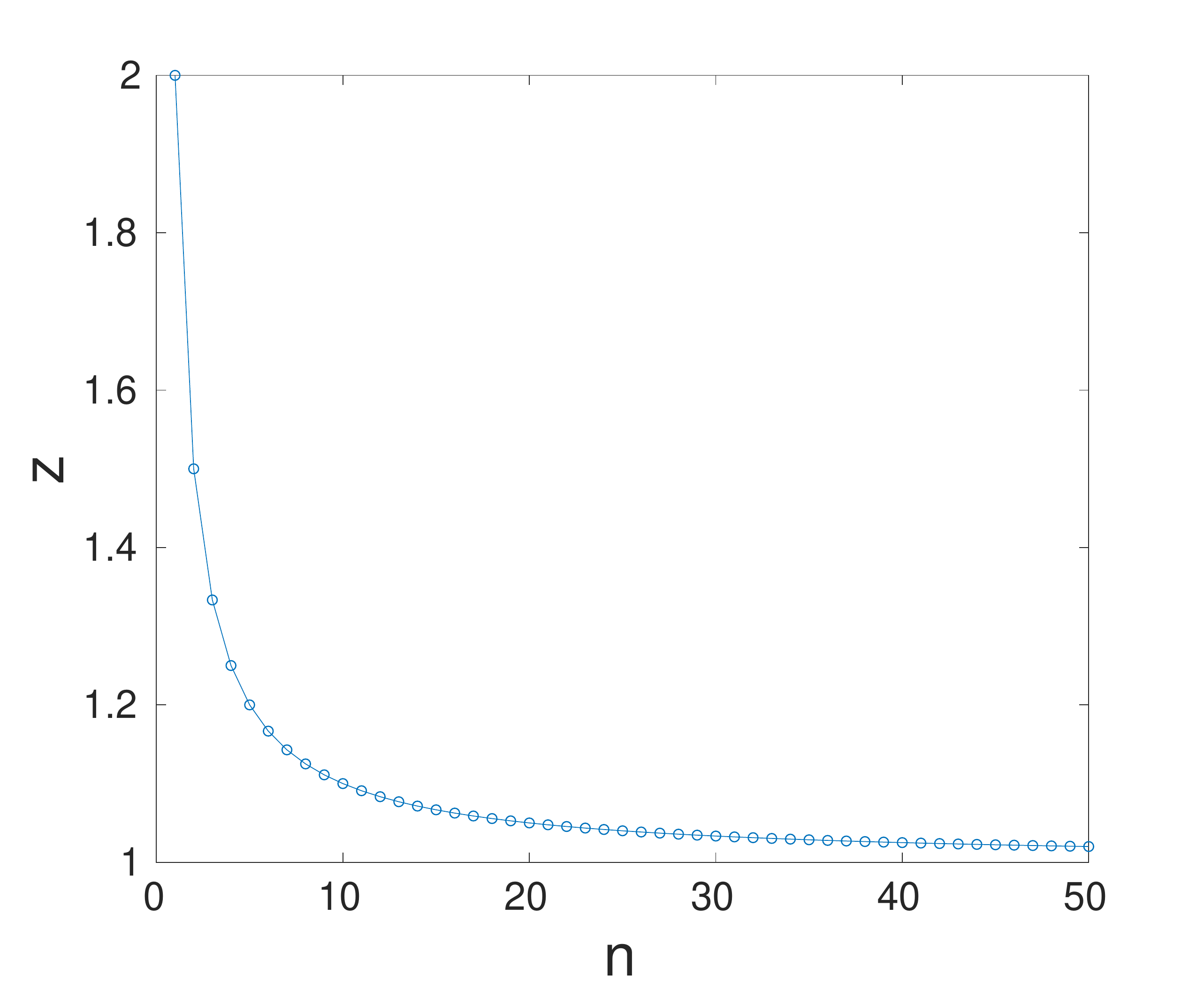}
}
\caption{(a)  Cell overlapping does not change the maximum neighbour number of six; thus, the underlying lattice is topologically triangular (yellow lines). (b) The average neighbour number versus layer number quickly drops to one. At $t=50$ days, $n\simeq 50$, $N\simeq 4000$ and $z=1.02$ which means $p_s=0.92$.   }\label{f overlap }
\end{figure}

In our construction there are no loops so that the deduction of the critical probability is similar to the one for a Bethe lattice or a tree graph. But, unlike regular lattices/graphs, in our model the neighbour number of a node is not the same for all nodes. To study our network we require the inclusion of  some less rigid regularities, the reason why we defined an {\bf average} neighbour number. In Fig. \ref{f overlap }(a) each cell is depicted by two concentric circles with a dark rigid core and  a lighter corona where overlap is allowed. Note that the maximum neighbour number is six no matter the extent of the corona. In our simulations a new cell must be in contact with another cells, but we allow cells to overlap to further study diffusion effects. Also, the random process of searching space for proliferation will slightly modify the geometry of the underlying network but not its topology, which will be the same as that of the triangular lattice, except for a few defects originated by a neighbor number lower than six. Thus density, which is relevant for the  diffusion of nutrient, oxygen or proteins for signalling,  does not play any role in the percolation problem.

\begin{figure}[h!]
\centering
\includegraphics[width=0.9 \linewidth]{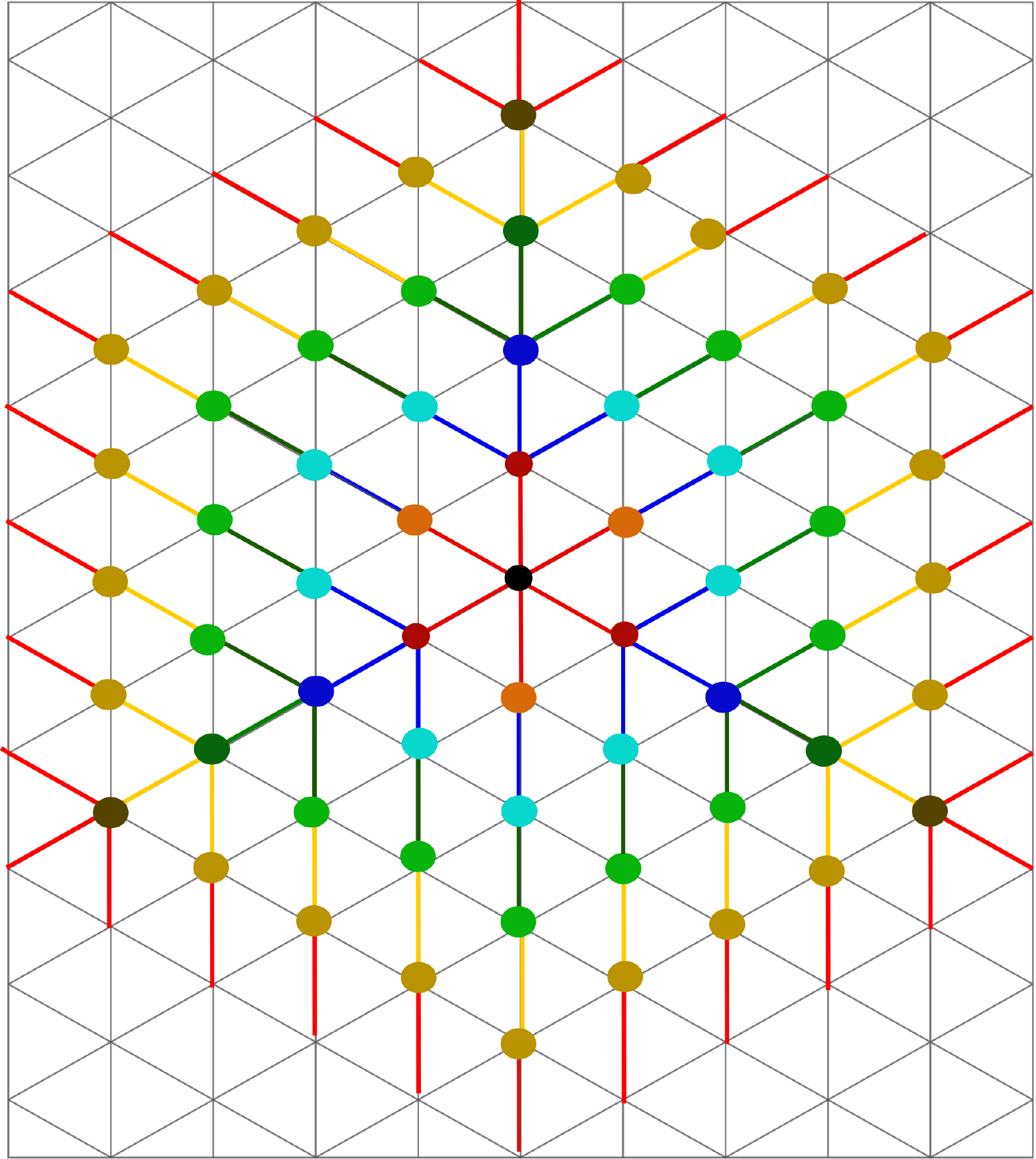}
\caption{The growth of the colony on a triangular lattice starting from the center (black dot). Each layer is depicted in a different color and contains three nodes with $z=3$ and the rest with $z=1$.     }\label{f bethetriangular }
\end{figure}

To roughly estimate $z$ we build up a tree graph over a triangular lattice substrate. A possible connection pattern of nodes in such a  network is depicted in Fig. \ref{f bethetriangular }, which was built layer  by layer with each layer represented in a different color. The layer $n=0$ is the central black dot that represents the seed. In this particular example, we depict in each layer, three nodes with  $z=3$ (darker dots) leaving the rest of the nodes with $z=1$ (lighter dots). It easy to see that, for any layer, there are $6\times n$  nodes for $n>0$ and, then, to connect  two layers without loops we need on average $z= \frac{n+1}{n}$. Because $1<z<2$ and quickly decays with $n$, c.f. Fig. \ref{f overlap }(b), we do expect that the critical probability for percolation, given by Eq. \eqref{e 1/z}, quickly shifts to $p_c=\frac{n}{n+1}\rightarrow 1$ as $n\rightarrow 1$. The  size of the network will be $N=6\sum_{i=1}^n i +1$  leading to the limit $p_c \xrightarrow[N \to \infty]{} 1$. Thus, as the colony increase its size, the percolation threshold shifts towards 1. The main consequence is that there is no chance to perform a finite size scaling to detect the critical transition point by simulations as in the case of 1D chains.  

As mentioned before, in Figs. \ref{f 2d examples} and \ref{f 2d examples2}, the proliferative cells are distributed in the last two outer layers. This feature of the colony occurs because the cells in the simulation are requested to proliferate in a random order which is also the cause of the observed circular shape of the colony. Because the vertices of the hexagonal array  shown in Fig. \ref{f bethetriangular } have a lower probability to be chosen for proliferation as system size increase, cells in their edges proliferate first, having at the beginning of each time step, more room to do so. Thus, the estimated number of cells in each layer is underestimated as well as their neighbour number. As a consequence, we will expect that the actual value of $p_c$ as a function of size or time will be much lower than that the predicted by Eq. \eqref{e 1/z} and our estimation of $z$. A precise estimation of $z$ will require a statistical treatment that is beyond of the scope of this article.

The result implies that the proper (mathematical) percolation transition will occur at $p_s=1$ as in 1D percolation. Such a result also implies that the only way for a CSC to be at the border, for an arbitrarily large colony, is when it is forced to non-differentiate. Nevertheless, monoclonal colonies cannot be maintained and grown forever. A 50-days experiment is very long-spanned and  rare to find in the literature. 

\begin{figure}[h!]
\centering
\includegraphics[width=0.9 \linewidth]{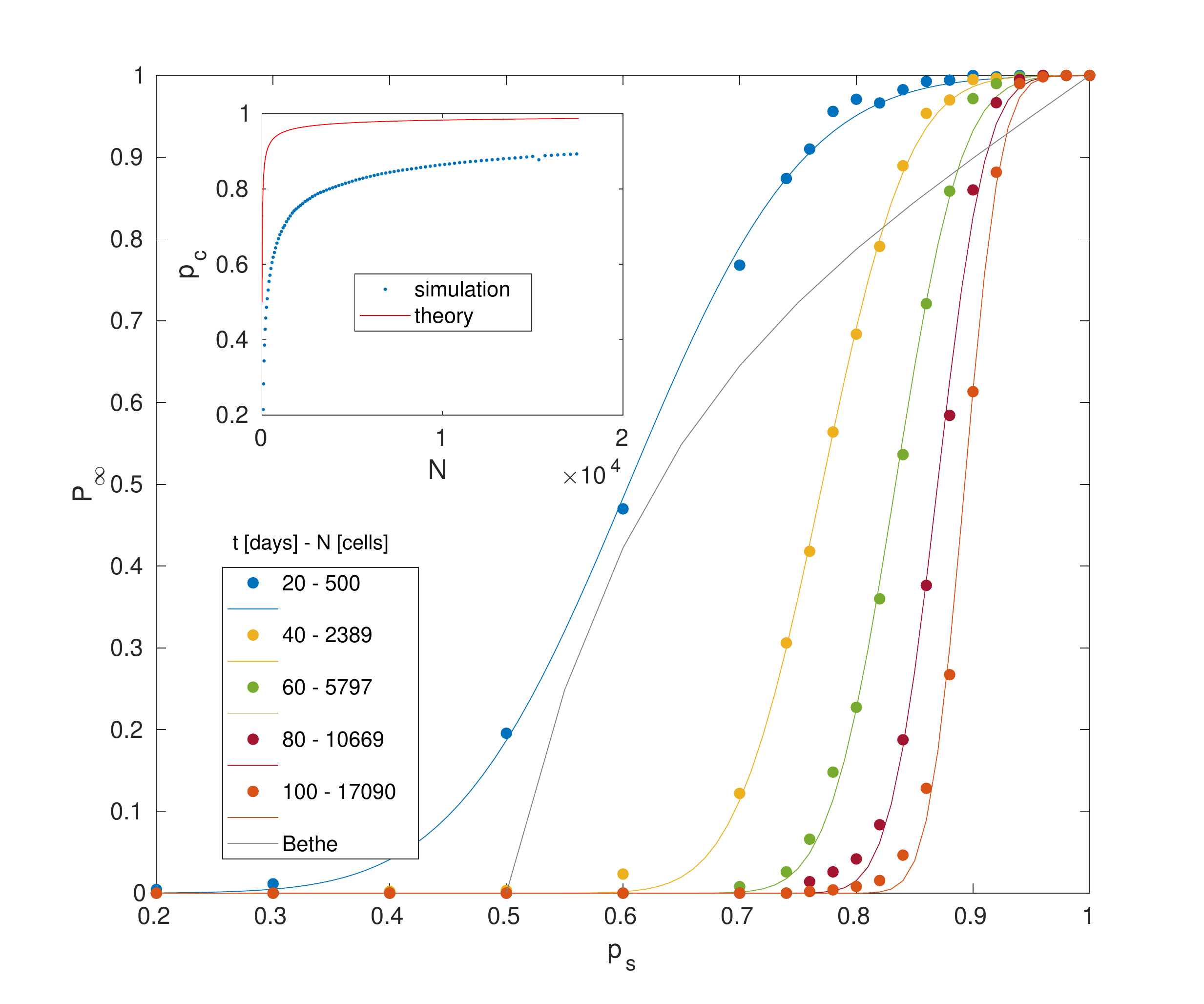}
\caption{Probability of percolation as a function of $p_s$ for  different colony sizes. Dots correspond to simulations and  lines are theoretical fittings to estimate the percolation threshold (see text). The gray line is the theoretical Bethe result.  Inset: Percolation threshold $p_c$ at different sizes measured by simulations (blue). Our simple theoretical approach overestimates this threshold(red) .   }\label{f percola2}
\end{figure}

The quantity that really matters is $P_\infty$, the probability for a CSC to be at the border, whose behavior is depicted in Fig. \ref{f percola2} as a function of $p_s$ at several times/sizes. Because the relationship between time and size is bijective, we report both values in the legend. In this graph, the dots are the result of the simulations presented in the previous section that measure the frequency of CSC in the border of the colony averaged over a thousand realizations. To obtain a suitable fitting of these data as continuous functions of $p_s$, we used erf functions following Yonezawa's classical work \cite{Yonezawa1989}. The inflection point of the erf curves coincides with the percolation threshold giving a good estimate of $p_c$. Besides, the theoretical result for the percolation parameter in a Bethe lattice of order 3 is depicted in gray for comparison purposes. Note that   $P_\infty$ does not jump as in Bethe percolation but the transition  is a continuous one as it is known for several regular lattices, including the triangular one.   Gonzalez \emph{el al.} \cite{Gonzalez2013} studied the transitions of several forms of percolation in triangular lattices; reporting the critical probabilities by finite size scaling analysis. They found universal features with values between 0.5 and 0.8 for the percolation threshold depending on the problem. On the contrary, and as expected, as our system size increases, the fitted curves become steeper and steeper and their inflection points predict a shift of $p_s$ towards 1 as depicted by the blue dots in the inset of Fig. \ref{f percola2}. Comparison with our theoretical result gives an overestimation of the percolation threshold, red line, as explained before. We can therefore expect that a CSC will be at the border of the colony in half of the realizations simulating a three weeks experiment.

\section{Conclusions}

Metastasis is an intriguing feature of cancer invasion. Most research in the field is guided by the biological point of view even if there are many mathematical models that attempt to describe   
it \cite{Iwata2000,Menchon2009,McGillen2014,Rhodes2019}.

In the examples of growth given in Fig. 1(a) or Fig. 2(a) it becomes clear that, under normal culture conditions, CSCs will remain in the inner core of spheroids. It is also deduced from the low probabilities $P_\infty$, shown in Fig. \ref{f percola2}, of finding a CSC at the border of the colony.  If this could be observed in the lab, the experiments that reveal a CSC preference for hypoxic environments can be easily explained by this fact. It was experimentally reported that hypoxia maintains the undifferentiated state of primary glioma cells, slow down the growth of glioma cells which were in a relatively quiescent stage, increases the colony-forming efficiency and migration of glioma cells, and elevates the expression of markers of stem cells, but the expression of markers for stem cell differentiation was reduced after hypoxia treatment \cite{Persano2011, Li2013}. It is also known that the inner core of spheroids and tumors has a lack of oxygen and nutrients that are first consumed by the outer layers of the tumor \cite{Liapis1982,Robinson1990,Jiang2005,Bull2020}. In this context, CSCs will be just a phenotype that has evolved to survive under hypoxic conditions in order to drive tumor growth.
 
On the other hand, metastasis requires CSCs to surpass three major barriers, the probability of an active CSC to be present at the border of the colony, the probability that it detaches from the tumor and the probability to find a suitable place to proliferate. The results of the present work establish that the first of those probabilities is really small even for relatively small tumors supporting the rare occurrence of metastasis.       

For large self-replication rates, possibly generated by the environmental conditions, there are many active CSCs that would become candidates to detach. At present, we do not know of any report of a high amount of CSCs in tumors \emph{in vivo}  or in transplanted xenographs under normal physiological conditions that experimentally support a large self-replication rate as a part of the metastasis mechanism.  

An intriguing option is the possibility that metastasis will begin at early stages of the tumor progression\cite{Gray2003,Hosseini2016,Linde2018}. This is deduced from Fig. \ref{f funciones de t}(a) where the number of active CSCs shows a peak for low $p_s$ and short times. If this is the actual situation,  a CSC  must detach from the primary tumor in its first week of life where the probability to be in the border is close to 1 even for low values of self-replication. Note that, at this time, the tumor has no more than a dozen cells. This fact leads to us to the hypothezise  that a primary tumor is indeed the metastasis of some small young tumors. In this hypothesis the primary tumor is indeed the first colony able to fix and grow, while the metastatic ones come from later CSCs that take more time to attach successfully to a new location. The detached cells must survive competing for nutrients and space with the primary spheroids, which, at this stage, are hard contenders \cite{Barberis2015}.

Our two-dimensional modelcan be extended to three dimensions. This would be more realistic and also computationally more expensive. Preliminary results are qualitatively similar to those reported here but, to the date, we lack the statistics to perform a quantitative comparison. Another feature, obtained by changing geometrical constraints in our code, are simulations of non-solid tumors such as hematologic neoplasm. Preliminary results show that active CSCs are in a larger fraction than in solid tumor spheroids. As a consequence, there is a larger chance in neoplasms for CSCs to detach and develop metastasis.

Summarizing, percolation provides a way to develop a geometrical theory to support or complement signalling pathways, quorum sensing an other tools frequently used to study metastasis. Further experimental research will allow to elucidate if CSCs are the survivors with greater fitness as suggested by our present results.

\section*{Acknowledgements}
This work was supported by SECyT-UNC (project 05/B457) and CONICET (PIP 11220150100644), Argentina. The author is grateful to Dr. C.A. Condat and Dr. L. Vellón for fruitful discussions.

\bibliographystyle{ieeetr} 
\bibliography{geo_biblio}

\end{document}